\documentclass[twocolumn,showpacs,preprintnumbers,amsmath,amssymb]{revtex4}

\usepackage{graphicx}
\usepackage{dcolumn}
\usepackage{bm}

\begin{document}

\newcommand{\goo}{\,\raisebox{-.5ex}{$\stackrel{>}{\scriptstyle\sim}$}\,}
\newcommand{\loo}{\,\raisebox{-.5ex}{$\stackrel{<}{\scriptstyle\sim}$}\,}

\title{Evolution of the statistical disintegration of finite nuclei toward 
high energy}

\author{A.S.~Botvina$^{1,2}$, N.~Buyukcizmeci$^{3}$, M.~Bleicher$^{1,2,4}$}

\affiliation {$^1$Institut f\"ur Theoretische Physik, J.W. Goethe 
University, D-60438 Frankfurt am Main, Germany} 


\affiliation{$^2$ Helmholtz Research Academy Hesse for FAIR (HFHF), 
GSI Helmholtz Center, Campus Frankfurt, Max-von-Laue-Str. 12, 
60438 Frankfurt am Main, Germany}

\affiliation{$^3$Department of Physics, Selcuk University, 42079 Kamp\"us, 
Konya, T\"urkiye}

\affiliation {$^4$GSI Helmholtz Center for Heavy Ion Research, 
Planckstr.1, Darmstadt, Germany} 



\date{\today}

\begin{abstract}

We develop a statistical approach for the description 
of complex nuclei formation from dynamically produced baryons in 
high energy heavy-ion reactions. 
We consider a finite highly-excited expanding 
nuclear system formed after central nucleus-nucleus collisions. This 
system is sub-divided into primary equilibrated nucleon clusters. 
The final nuclei are produced after the decay of these excited clusters. 
By the successful comparison with the FOPI experimental data 
we prove the possibility of such a local 
equilibrium in nuclear matter with the temperature corresponding 
to the phase coexistence region. The regularities obtained in this 
new nuclei production mechanism are shown.

\end{abstract}


\maketitle


\section{Introduction}

Statistical models have a long and very successful history in nuclear
physics. Most prominently they have been used for the description of 
nuclear decay when an equilibrated 
source can be identified in the reaction. The most famous example of such a
source is the 'compound nucleus' introduced by Niels Bohr in 1936 
\cite{Bohr}. Such a compound structure was clearly seen in low-energy 
nuclear reactions leading to 
excitation energies of a few tens of MeV. It is remarkable that this 
concept is also applicable for nuclear reactions induced by particles 
and ions of intermediate and high energies, when nuclei break-up into many 
fragments (multifragmentation) \cite{SMM,Gross}. In fact a large amount of 
experimental data concerning multifragmentation reactions were successfully 
described in this way 
\cite{ALADIN,Bot92,MMMC,Bot95,Xi97,Ogu11,EOS,Vio01,Pie02,FASA,FASA2,MSU,INDRA,
TAMU,Dag2}. 
Generally, this process 
is associated with the manifestation of the nuclear liquid-gas 
type phase transition at subnuclear densities. 
In these reactions one can extend the statistical approach towards 
the finite systems, 
and demonstrate that it works when these systems decay rapidly. 
As was found in previous theoretical analyses the system excitation energy 
reaches about 10 MeV per nucleon, 
and one can consistently describe all experimental data 
with relatively low temperatures (T$\loo$6--8 MeV) of the thermal sources. 
At very high excitation energies observed in 
finite systems there were usually problems to describe the kinetic energies of 
produced fragments. Phenomenologically, this was resolved by introducing the 
regular (hydrodynamical-like) flows, and by assuming only chemical equilibrium 
in the systems \cite{Neu03}. 
In this paper we propose a novel method to extend the statistical 
approach to highly excited finite nuclear systems. To this aim 
we consider local statistical equilibrium for 
the complex nuclei formation in separate parts (clusters) of nuclear matter. 
We show that our method is fully consistent with all observables concerning 
the production of nuclei. Therefore, it provides new 
insight on the fragment formation in high energy reactions.

\section{Formation of excited nuclear statistical systems}

According to the statistical hypothesis, the initial dynamical 
interactions between nucleons lead to a re-distribution of the available 
energy among many degrees of freedom, therefore, the nuclear system 
evolves towards equilibrium. 
In the most general consideration the reaction may be 
subdivided into several stages: (1) a dynamical stage leading to 
formation of an equilibrated nuclear system, (2) the statistical fragmentation 
of the system into individual primary fragments, 
which can be accompanied by the deexcitation of hot primary 
fragments if they are in the excited states. 
Many transport models are used for the description of the dynamical 
stage of the nuclear reaction at high energies. They take into account 
the hadron-hadron interactions including the secondary interactions and 
the decay of hadron resonances. For this reason they may preserve some 
correlations between hadrons originated from the primary interactions in 
each event, which are ignored when we consider the final inclusive 
particle spectra only. Within dynamical models it is established that 
many particles are involved in this process via the intensive rescattering 
and the collective interaction during the primary collisions. In peripheral 
collisions the produced high energy particles leave the system and the 
remaining nucleons form an excited system (a residue). This system can 
collectively expand, and the residual interaction drives it towards 
an equilibrium at a low density. We may expect that the system evolves 
toward a state which is mostly determined by 
the statistical properties of the excited nuclear matter. Generally 
this equilibration may not be a complete one. However, as shown by comparison 
with experiment, in many cases it is sufficient to apply statistical 
theory for the nuclear fragment formation. Usually, the statistical approach 
has been applied to the excited residual nuclei formed from the spectator 
parts of 
the colliding nuclei, as well as for a nuclear system produced after their 
fusion (partial or full one). It was treated via the compound nucleus 
concept at low excitation energy \cite{Bohr}. This concept was 
generalized to the multifragmentation of a single nuclear 
source with a high energy which is sufficient for its fast thermal expansion 
before the disintegration \cite{SMM,Gross,FASA}. 

However, there is another interesting possibility for the application of 
the statistical approach to describe 
the evolution of the produced diluted nuclear matter: 
At the end of the dynamical stage (at a time around $\sim$10-30 fm/c after the 
beginning of the nucleus-nucleus collision) many new-born baryons and nucleons 
are escaped from the colliding nuclei remnants. Some of these baryons may be 
located in the vicinity of each other with local subnuclear densities 
around $\sim$0.1$\rho_0$ ($\rho_0 \approx 0.15$ fm$^{-3}$ being the ground 
state nuclear density). This nuclear matter density is very similar to 
the densities expected in the freeze-out volume which is assumed in the 
statistical approach as the proper place of the nuclei formation. Namely 
at this place the interaction between nucleon can still lead to fragment 
formation from these nucleons. Such nucleation processes will be improbable 
both at too low and too high densities. However, the system has to pass the 
above-mentioned density during its expansion, which allows to use 
the statistical models at this local space-time region. 

To investigate this process, in the first approximation, it is instructive to 
consider a general situation of baryonic nuclear matter expanding as a result 
of the previous dynamical process within a simple controlled model. 
For example, we can simulate an expanded nuclear matter state with 
stochastically distributed baryons.
We call our first method the phase space generation (PSG) method: Here, we 
perform an isotropic generation of all baryons of the excited nuclear system 
according to the microcanonical momentum phase space distribution with 
total momentum and energy conservation. It is assumed in the one-particle 
approximation that all particles are in a large volume (at subnuclear 
densities) where they can still interact with others to populate the phase 
space uniformly.  Technically, this is done using the Monte-Carlo method 
applied previously in the microcanonical SMM and Fermi-break-up model 
\cite{SMM}, and taking into account the relativistic effects according to 
the relativistic connection between momentum $\vec{p}$, mass $m$, and kinetic 
energy of particles $E_{0}$, see Eq.~(\ref{relc}). In Eq.~(\ref{relc}) the sum 
is over all particles and we use units with $\hbar = c = 1$. 
\begin{equation} \label{relc}
\sum \sqrt {\vec{p}^2 + m^2} = E_0 + \sum m .
\end{equation}
The total kinetic energy available for the motion of baryons 
$E_{0}$ (we call it the source energy) is the important parameter 
which can be adjusted to describe the energy accumulated in the system 
after the dynamical stage. We believe that the PSG method is a reasonable 
assumption due to the very intensive interactions between the colliding 
nucleons of the target and projectile, which take place in 
some extended volume during the reaction, leading to the equilibration 
of the one-particle degrees of freedom. Note, that this is not an equilibrium 
with respect to the nucleation process. In this case we do not take directly 
into account the coordinates of the baryons but we assume they are 
proportional to their velocities and strictly correlate with them. This is 
also consistent with the results of dynamical models. 

In the second method,  we assume the momentum generation similar to 
the explosive hydrodynamical process when all nucleons fly out from the 
center of the system with the velocities exactly proportional to their 
coordinate distance to the center of mass.  We call it the hydrodynamic 
generation (HYG) method. In this method we pace randomly (Monte-Carlo) 
all nucleons uniformly inside a sphere with the radius 
$F\cdot R_n \cdot A_0^{1/3}$ without overlapping. 
Here $A_0$ is the nucleon number, and $R_n \approx 1.2$~fm is the nucleon 
radius. The scaling factor $F \approx 3$ is assumed to describe the expanded 
volume in which the nucleon can still strongly interact with each other. 
At intermediate collision energies this volume corresponds approximately 
to the average expansion of the system in line with the 
transport model simulations, when the baryon interaction rate drastically decreases. 
Finally, we attribute 
to each nucleon a velocity by taking into account the momentum and energy 
conservation for the relativistic case (Eq.~\ref{relc}). 
Obviously, the velocities and coordinates of baryons are strongly 
correlated with each other. 

\begin{figure}[tbh] 
\includegraphics[width=8.5cm,height=13cm]{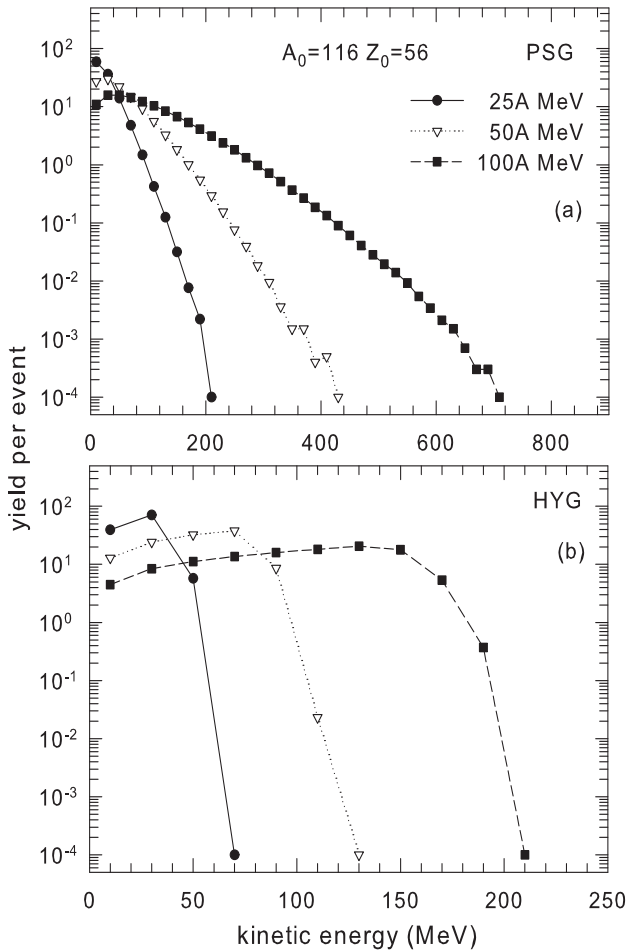} 
\caption{\small{ 
Energy spectra for initial nucleons of the hot expanding nuclear system 
according to the microcanonical phase space distribution - PSG (top panel (a)) 
and according to the hydrodynamical-like explosion - HYG (bottom panel (b)). 
The assumed total kinetic energies are 25, 50, and 100 MeV per nucleon. 
The nucleon source size and composition are shown in the top panel. 
}} 
\label{fig1} 
\end{figure}

For illustration, in Fig.~\ref{fig1} we demonstrate the energy distribution 
of nucleons generated using the PSG and HYG methods. Here we assume an 
intermediate source with $A_0$=116 and charge $Z_0$=56. However, we 
have found that the general trends do not depend on the system sizes. 

It is obvious that both the PSG and HYG methods generate the baryonic 
matter which expands in each coordinate point. All parts of this matter 
do certainly pass through the 'freeze-out' density where nuclei can be 
still formed as supposed in the statistical models. However, because of 
the different momenta and locations of the nucleons, the different 
parts of the system pass this density at different times. 
Therefore, one may not claim that the whole nuclear system is in the 
same statistical freeze-out volume concerning the nuclei formation. 
However, it is possible to assume local equilibrium. 
Let us stress that it is important to consider the PSG and HYG methods 
as complementary descriptions of the finite expanding 
system, corresponding to different limits of the dynamical description.

\section{Subdivision of the nuclear system into excited clusters}

The idea is to divide the low-density nucleon matter into small parts 
(clusters) with nucleons which are in equilibrium respective to the nucleation 
process \cite{Bot21}. These clusters are analogous to the local freeze-out 
states for the liquid-gas type phase coexistence adopted in statistical 
models. Since the nucleons are moving with respect to each other these clusters 
are excited objects and distributed over the whole nuclear system. 
The subsequent evolution of 
such clusters, including the formation of nuclei from these baryons, can be 
described in the statistical way. Within our procedure we can claim that 
these hot clusters decay into nuclei. It may look like that the formation 
of the clusters is similar to the standard coalescence procedure \cite{Ton83}. 
However, it is assumed in the simplistic coalescence picture that only 
nucleons which combine a bound nucleus can interact in the final state. 
All other nucleons will not interact with this nucleus, or interact very 
slightly by taking extra energy to conserve the momentum/energy balance. 
In our case all nucleons of the primary clusters are fully involved in the 
interaction leading to final nuclei. 
Nevertheless, not all of these nucleons will be bound in the nuclei in the end. 

Since the matter expands, a crucial question is, if the interactions between 
the baryons inside these clusters are sufficiently strong to lead to local 
equilibrium. In this case they could be considered as statistical subsystems 
where the phase-space dominates the nuclei formation. We remind the reader 
that the lifetime of finite nuclear species is related to the 
energy accumulated into these species. We know from the extensive studies 
of nuclear multifragmentation reactions \cite{Bot95,Xi97,Ogu11,SMM,Bot92,MMMC} 
that the excitation energies of the excited nuclear residue systems can 
reach up to 8--10 MeV per nucleon, and the statistical models describe their 
disintegration very well. We have also learned from the analysis of nuclei 
production in multifragmentation that the densities before the break-up of 
these systems are around 0.1--0.3$\rho_0$, and their lifetime is 
50--100 fm/c \cite{Vio01,FASA}. 
We suggest that the difference between  the multifragmentation of excited 
projectile- and target--like residues and the formation of the baryon clusters 
in the expanded matter is just due to the dynamical mechanisms leading to 
these diluted finite systems. In the spectator multifragmentation the systems 
are prepared after the dynamical knocked-out of many nucleons and thermal 
(or dynamical) expansion of the remaining nuclei. Our new baryon clusters 
can be formed 
as a result of the local interaction of the stochastically produced primary 
baryons. Therefore, we can estimate that an energy around $\sim$10 MeV per 
nucleon is a reasonable value which can be reached in such hot 
stochastic clusters, similar to the standard multifragmentation case. 
If the excitation energy is much higher, then the existence of such clusters 
as intermediate finite systems, including their following evolution in the 
statistical way, become problematic. We think that the final conclusion on 
the excitation energy can only be done after a detailed comparison with 
experimental data. 

To describe the cluster formation we suggest the coalescence prescription, 
and apply the coalescence of baryon (CB) model \cite{neu00,Bot15}. 
In the PSG and HYG cases the coalescence criterion is the proximity of the 
velocities (or momenta) of the nucleons. In the both cases we do not need to 
include explicitly the coordinate of nucleons, because in the PSG and HYG 
approaches the velocities and space coordinates are already correlated. In 
particular, the coordinate vectors are directly proportional to the velocities 
vectors. So the velocity coalescence parameter is sufficient for the 
cluster identification in these models. Such a strong space-momentum 
correlation exists in 
many explosive processes and it influences the original clusterization. 
However, for the following evaluation of the cluster properties we assume 
that such clusters with nucleons inside have the density of 
$\rho_c \approx \frac{1}{6} \rho_0$ as it was established in the previous 
studies of statistical multifragmentation process \cite{SMM,Gross,Vio01,FASA}. 
This corresponds to the average distance of around 2 fm between neighbour 
nucleons, and the interaction between these nucleons can lead to the 
formation of nuclei. 
Within the CB model we assume that baryons (both nucleons and hyperons) can 
produce a cluster with mass number $A$ if their velocities relative to the 
center-of-mass velocity of the cluster is less than $v_c$. Accordingly we 
require $|\vec{v}_{i}-\vec{v}_{cm}|<v_{c}$ for all $i=1,...,A$, where 
$\vec{v}_{cm}=\frac{1}{E_A}\sum_{i=1}^{A}\vec{p}_{i}$ ($\vec{p}_{i}$ are 
momenta and $E_A$ is the sum energy of the baryons in the cluster). 
This is evaluated by sequential comparison of the velocities of all baryons. 
To avoid problems related to the sequence of nucleons within the algorithm, 
we apply the iterative coalescence procedure \cite{neu00,Bot15}, starting 
from a small coalescence parameters for clusters 
and increasing it step-by-step up to $v_c$. 

\begin{figure}[tbh]
\includegraphics[width=9cm,height=16cm]{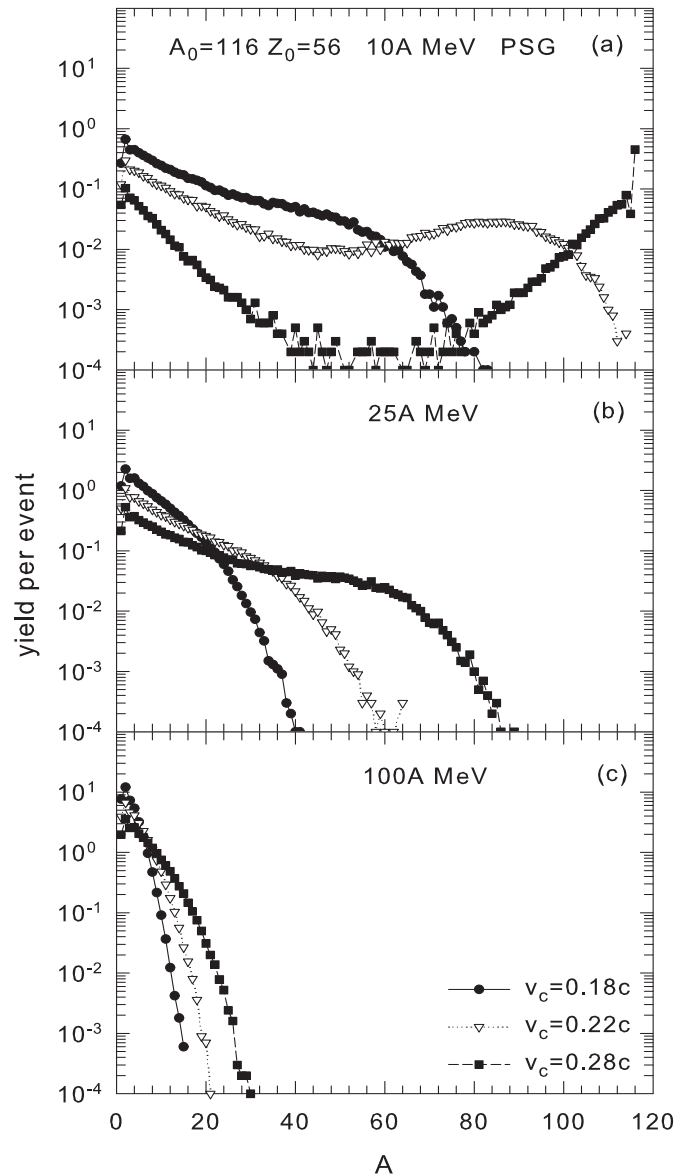} 
\caption{\small{ 
Yield of coalescent-like clusters versus their mass number A after the CB 
calculations at the source energy of 10, 25 and 100 Nev per nucleon. 
Composition and sizes of sources, nucleon generator (PSG), as well 
as coalescence parameters ($v_c$) are indicated in the panels. 
}}
\label{fig2}
\end{figure}

We show in Fig.~\ref{fig2} the distributions of clusters as a function of 
mass number 
$A$ after the coalescence of initial nucleons of the primary source $A_0$=116, 
$Z_0$=56, for $E_{0}$=10 A MeV (top panel), $E_{0}$=25 A MeV (middle panel), 
and $E_{0}$=100 A MeV (bottom panel), for the velocity coalescence parameters 
$v_c$=0.18, 0.22, and 0.28 $c$. We only show the results for the PSG model 
since the HYG model leads to a qualitatively similar picture. In our 
case $v_c$ is the maximum velocity deviation and all baryons with lower 
relative velocities do compose a cluster. The middle $v_c$=0.22 $c$ is 
approximately of the order of the Fermi-velocity which is expected in such 
nuclei. It is obvious that the smaller $v_c$ will lead to small clusters 
without excitation energies and these values are 
consistent with the coalescence parameters 
extracted from the analyses of experimental data for the production 
of lightest nuclei in previous years \cite{Ton83,Bot17a}. 
One can see that the large primary clusters can indeed be produced within this 
mechanism even at the high source energy. At low energies we are naturally 
transiting to the compound-like state with one big excited nucleus. 

Such a clusterization procedure is fully consistent with the consideration 
of the low-energy compound nucleus processes and the 
multifragmentation processes which take place in one source at low and 
moderate excitation energies. 
Apparently, in our case we shall deal with several local statistical sources 
(clusters) in one events. This procedure can also be suggested as a 
generalization for the statistical description of the 
disintegration of highly excited finite nuclear systems which 
are produced in intermediate and high energy nucleus collisions.

\section{Disintegration of excited clusters into nuclei} 

The excited primary nuclear clusters will disintegrate into small peaces. 
As mentioned above, this disintegration can be considered as a result of the 
residual nuclear interaction at the subnuclear density  between baryons 
of these clusters leading to the formation of final nuclear species. 
In the end the cold and stable nuclei are produced. 
The energy accumulated in such low-density finite clusters 
is the main ingredient which determines their following evolution. 
In the lowest limit we 
can estimate this excitation as a relative motion of the nucleons initially 
captured into a cluster respective to the center of mass of this cluster. 
In this case the excitation energy $E^{*}$ of the clusters ($j$) 
with mass number $A$ and charge $Z$ is calculated as 
\begin{equation} \label{excit}
E^{*}= \sum_{i=1}^{A}\sqrt{\vec{p}_{r i}^{ 2}+m_{i}^{2}} - M_{A} ,
\end{equation}
where $M_{A}$ is the sum of the masses of the nucleons in this nuclear cluster, 
$i=1,...,A$ enumerates the nucleons in the cluster, 
$m_{i}$ are the masses of the individual nucleons in the cluster, 
$\vec{p}_{r i}$
are their relative momenta (respective to the center of mass of the cluster). 
However, in the cluster volume the nucleons can interact with each other 
and the binding interaction energy $\delta E^{*}$ should be added to 
the $E^{*}$. As an upper limit we can take the ground state binding energy 
of normal nuclei with $A$ and $Z$. However, 
since our clusters present pieces of nuclear matter expanded 
already during the previous dynamical reaction stage, 
we suggest that this energy should be lower. 
Therefore, as first approximation we use the following recipe for 
the evaluation of $\delta E^{*}$: It is 
known the ground state binding energy of nuclei can be written as the 
sum of short range contributions ($E_{sr}$, which naturally includes volume, 
symmetry, surface energies), and the long-range Coulomb energy ($E_{col}$), 
see, e.g., Ref.~\cite{SMM}. Since a cluster is extended, its Coulomb 
energy contribution will be smaller and we can recalculate it proportional 
to $\left(\frac{\rho_c}{\rho_0}\right)^{1/3}$ (in the Wigner-Seitz 
approximation \cite{SMM}). For the short range energies, it is assumed that 
all contributions do also decrease proportional to 
$\left( \frac{\rho_c}{\rho_0} \right)^{2/3}$ as it follows 
from the decreasing of the Fermi energy of nuclear systems. 
This prescription can be used for any description of the initial dynamical 
expansion of the system, for example, with the transport models. 
This was worked out in our previous paper, Ref.~\cite{Bot21}, 
where we have suggested 
\begin{equation} \label{deltaE}
\delta E^{*}= E_{col}\left(\frac{\rho_c}{\rho_0}\right)^{1/3} + 
E_{sr}\left(\frac{\rho_c}{\rho_0}\right)^{2/3}~. 
\end{equation}
It provides a reasonable estimate in between the mentioned limits. 
In the following we  call this energy the cluster excitation energy, 
or the cluster internal energy. 

In the present work we use only the PSG and HYG methods to generate nucleon 
distributions. In this 
case we are able to take into account the energy conservation in the expanded 
system. 
Since after the subdivision of nucleons we consider the isolated hot clusters 
it is reasonably to scale the cluster's internal excitation to fit the 
total energy 
balance. After the nucleon momentum generation the excitation of clusters is 
the only quantity which can correctly be used for this purpose. Therefore, 
finally for the cluster $j$ we take 
\begin{equation} \label{excit2}
E^{*}_{j}= \beta \cdot (E^{*}+\delta E^{*}) ,
\end{equation}
where $\beta$ is found from the energy balance in the system: 
\begin{equation} \label{excit3}
\sum_{j=1}^{N}(\sqrt{\vec{p}_{j}^{ 2}+m_{j}^{2}} + E^{*}_{j}) =  E_t + M_t. 
\end{equation}
Here, $N$ is the number of clusters in the system, $\vec{p}_{j}$ and $m_{j}$ 
the cluster momenta and masses. The right part contains the initial total 
energy $E_t$ deposited into the system and the initial mass of the system $M_t$. 
Further we take this new excitation energy $E^{*}_{j}$ for the statistical 
calculations of the nucleation process, i.e., for the decays of all excited 
clusters. 

\begin{figure}[tbh]
\includegraphics[width=8.5cm,height=13cm]{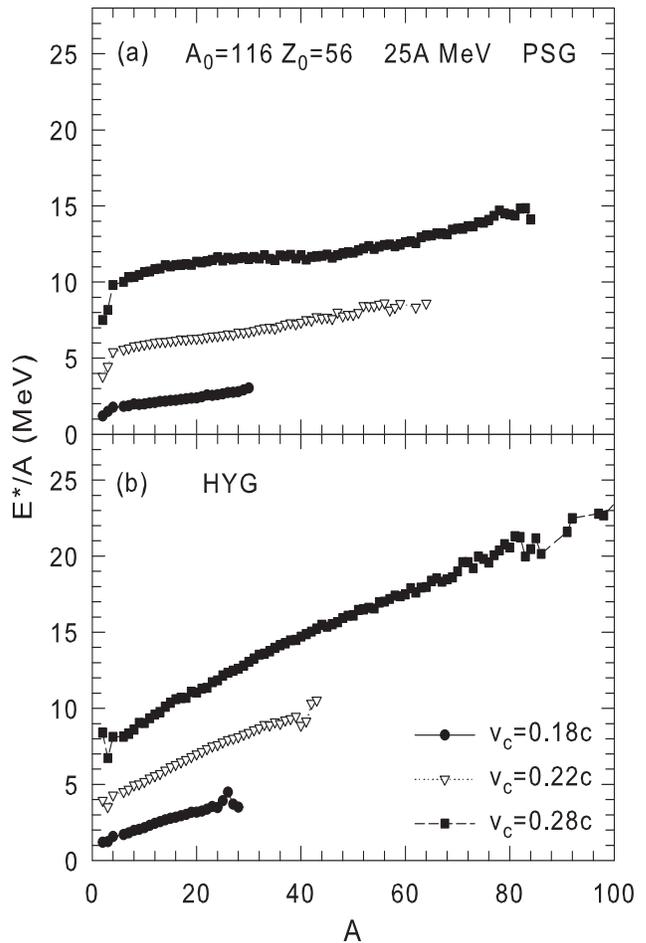}
\caption{\small{ 
Average internal energy of coalescent clusters versus their mass 
number $A$ produced as a result of 
the coalescence (CB) in the sources with $A_0$=116 and $Z_0$=56 
after PSG (top panel (a)) and HYG (bottom panel (b)). 
The source energy and coalescence parameters are shown in the panels. 
}}
\label{fig3}
\end{figure}

For illustration, in  Fig.~\ref{fig3} 
we present the average internal energies of such clusters versus their 
mass number for the big systems $A_0$=116, $Z_0$=56, and $E_{0}$=$25A$ MeV, 
with the 
coalescence parameters $v_c$ from 0.18, to 0.28 $c$. One can see that 
the internal energy per nucleon increases with the parameter $v_c$. This is 
because more nucleons with large relative velocities are captured into the 
same cluster. By comparing the panels of Fig.~\ref{fig3} we see the effect 
of the source generator on these distributions: The internal energies are not 
very different, since 
they are determined by the relative nucleon motion inside the clusters. 
Nevertheless the HYG provides a general increase of the internal energy with 
the mass number since the large clusters are consisting of baryons 
having initially higher velocities. 

As was done previously in the analyses of heavy-ion collisions at low and 
intermediate energies we use the statistical 
multifragmentation model (SMM) \cite{SMM} to describe the break-up of normal 
nuclear clusters. This approach includes (consistently connected) 
multifragmentation, evaporation, fission (for large nuclear systems), and 
Fermi-break-up (for small systems) models. Therefore, it can be used as 
an universal model to describe the decay of single statistical sources 
from very low to rather high excitation energies. At the same time it reflects 
properties of nuclear matter resulting in a phase transition. We remind 
the reader that this approach is very successful in the description of the 
disintegration of highly excited nuclear systems, as was demonstrated by 
numerous comparisons with multifragmentation data in peripheral 
relativistic collisions and in central nucleus-nucleus 
collisions around the Fermi energy 
\cite{ALADIN,Xi97,Ogu11,EOS,Pie02,FASA,FASA2,MSU,INDRA,Dag2}. 

\begin{figure}[tbh]
\includegraphics[width=8.5cm,height=16cm]{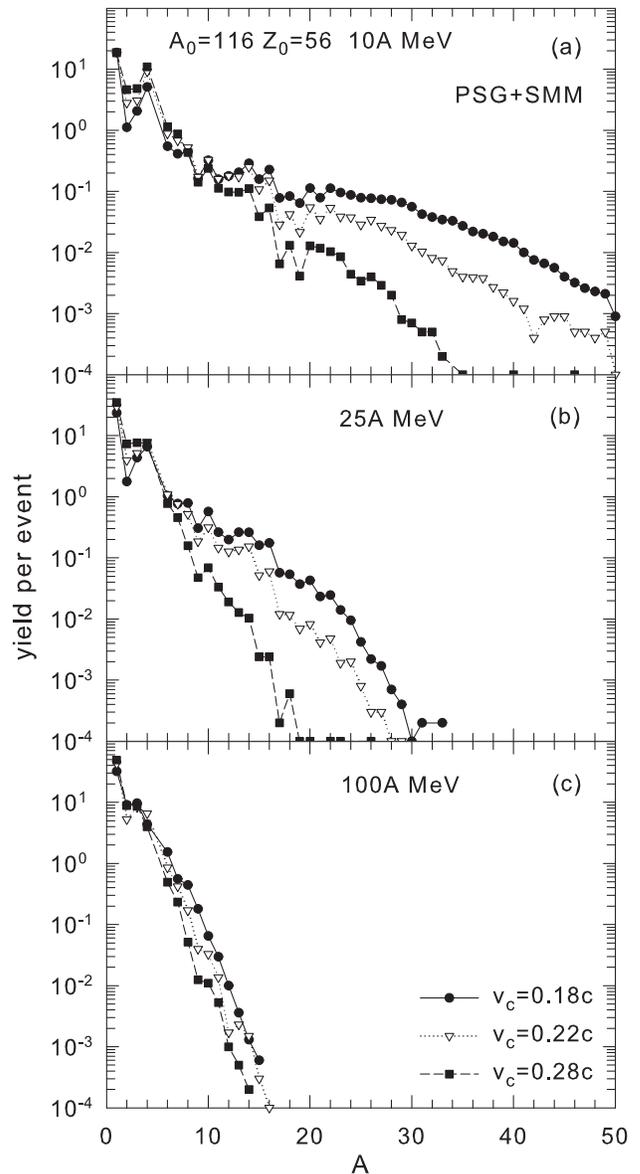}
\caption{\small{ 
Yield of final nuclei versus their mass number A after the de-excitation 
of primary clusters shown in Fig.~\ref{fig2}. The notations are as in 
Fig.~\ref{fig2}. 
}}
\label{fig4}
\end{figure}

In Fig.~\ref{fig4} we demonstrate the fragment yields obtained after the 
de-excitation of primary clusters. We have taken the same cases as 
were shown in Fig.~\ref{fig2}. 
We obtain a well known general regularity that at high initial excitation 
energy the yield of nuclei decreases exponentially with their masses. 
And it becomes even steeper at the highest source energies. One may naively 
conclude from this observation that the source's temperature becomes 
higher. However, in the case of finite systems the situation is 
different. Actually, two effects contribute to this kind of 
behaviour: 1) The cluster's de-excitation leads to small fragments, and 
2) the size of the clusters 
decreases with increasing the initial energy, because the clusters can 
accumulate the limited amount of energy and the temperature of clusters 
may not change. It is 
interesting, however, that after this decay a small $v_c$ may provide even 
larger fragment yields than a larger $v_c$. Because in our cases of 
relatively low initial source energy, a small $v_c$ can still lead to the 
formation of sufficiently big clusters but with smaller cluster excitation 
energies. 

\begin{figure}[tbh]
\includegraphics[width=8.5cm,height=16cm]{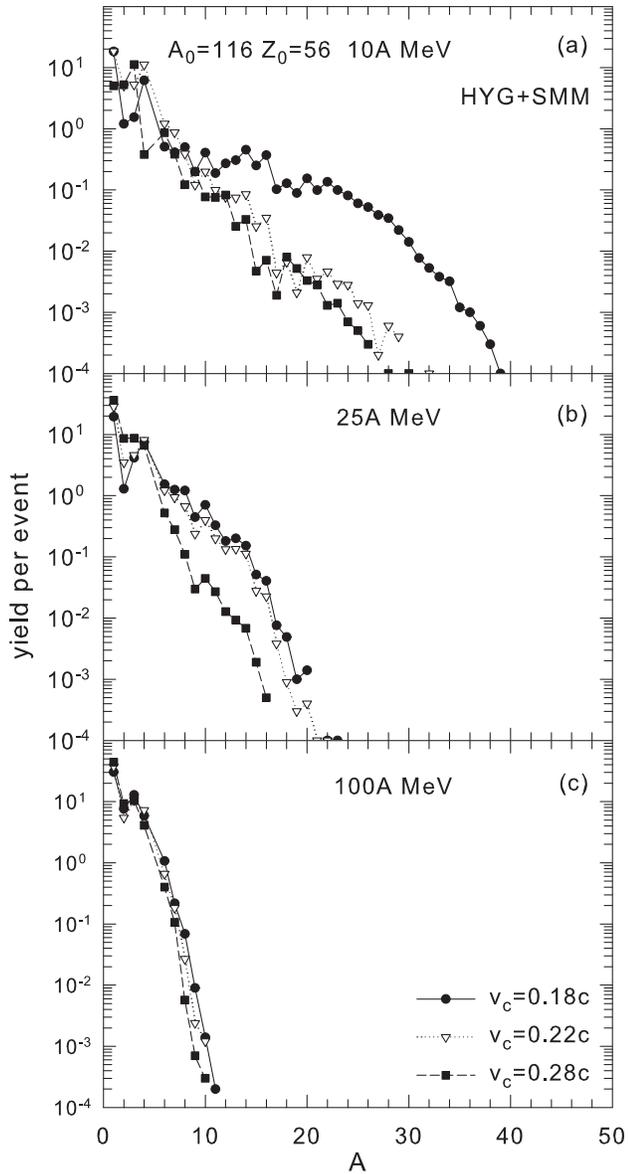}
\caption{\small{ 
Yield of final nuclei versus their mass number A after the de-excitation 
of primary clusters, by using the HYG initial nucleon distributions. 
Other notations are as in Fig.~\ref{fig4}. 
}}
\label{fig5}
\end{figure}

To investigate the influence of the initial nucleon distribution on the 
final fragment yields we show in Fig.~\ref{fig5} the calculations as in 
Fig.~\ref{fig4}, however, using the HYG nucleon generation method. We see 
qualitatively similar results concerning the yield evolution despite very 
different initial 
distributions (see Fig.~\ref{fig1}). This gives us some confidence that the 
conclusions presented here are robust and will also not change, if more 
elaborated initial calculations, e.g., from transport simulations are used. 

\begin{figure}[tbh]
\includegraphics[width=8.5cm,height=13cm]{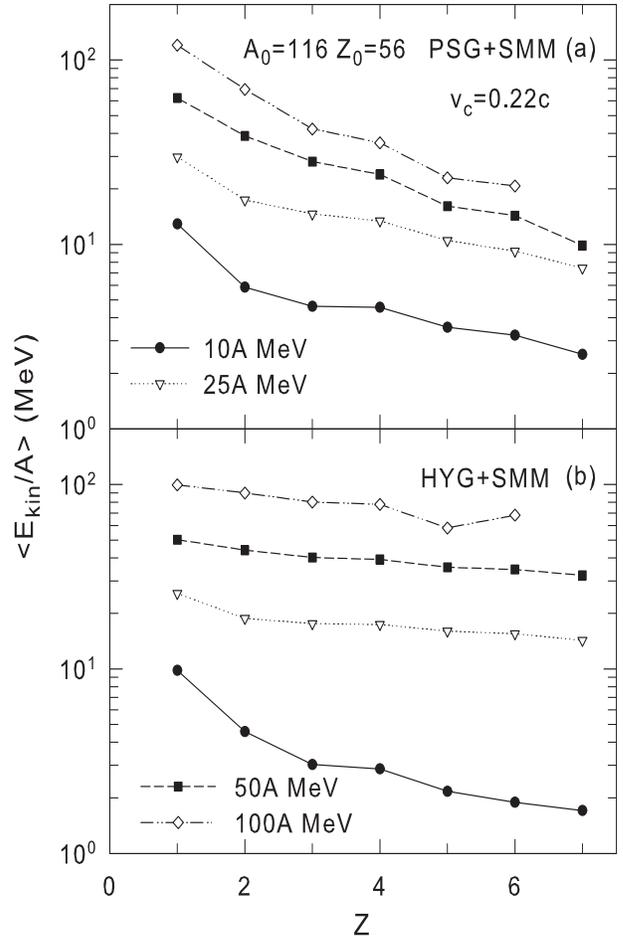}
\caption{\small{ 
Average kinetic energies (per nucleon) of final nuclei versus their charges 
$Z$. The calculations are performed for the initial source with $A_0$=116 
and $Z_0$=56 at the initial excitation energies of 10, 25, 50, and 100 MeV 
per nucleon. Notations with the parameters are in the figure. Top panel (a) 
is for the PSG initial nucleon generation, and bottom panel (b) is for the 
HYG one. 
}}
\label{fig6}
\end{figure}

Besides the yields, the kinetic energy of produced nuclei is also 
a very important characteristic of the nucleation process. In Fig.~\ref{fig6} 
we demonstrate the average kinetic energy of the final nuclei for the same 
initial nuclear system in the cases of PSG and HYG nucleon generations. 
Usually, the kinetic energy is associated with the nuclear fragment flow. 
For clarity we use only one coalescence parameter $v_c=0.22c$ for the 
primary cluster formation. 
One can see that substantial flows can be reached in those cases. Especially 
when 
we use HYG primary nucleons (bottom panel) at a high initial source energy. It 
is an obvious result, since many nucleons are concentrated at the high kinetic 
energy region in this case (see Fig.~\ref{fig1}). A moderate decreasing of the 
flow energy with the nuclei charges (consequently, with mass numbers) in the 
PSG case (top 
panel) is also understandable, since most nucleons have low kinetic energies 
according to the PSG generation method.

\section{Comparison with experimental data} 

In the previous section we have investigated the general regularities for the 
production of fragments which can be obtained within the proposed mechanism. 
Below we demonstrate the results of our hybrid approach which include the 
initial generation of the nucleon momenta (PSG and HYG), the selection 
of the primary excited clusters (CB), and the statistical description of 
the nucleation inside the clusters (SMM). We provide a comparison 
with the extensive high-quality experimental data obtained by the FOPI 
collaboration \cite{FOPI10}. 
They are obtained in Au+Au and Ni+Ni central collisions. 
Previously these fragment production data could not be consistently analyzed 
with neither dynamical models nor with statistical ones. 

One remark on the experimental data obtained by different groups 
should be made. The published data depends essentially on the selection of 
the central events. In particular, the FOPI group has used the ERAT criterion 
which includes the ratio of total transverse to the longitudinal kinetic 
energies of particles in the center-of-mass (c.m.) system \cite{FOPI10,FOPI97}. 
While in Ref.~\cite{Kunde} a simple criterion related to the light particle 
multiplicity is employed. As a result the extracted yields of the nuclei are 
slightly different. For example, the ratio of the intermediate mass fragment 
(with $Z \geq 3$) yields to the yield of $Z=3$ fragments in Au + Au central 
collisions at 100 A MeV/nucleon obtained in Ref.~\cite{Kunde} is nearly the 
same as the one obtained in the FOPI experiment, but at 120 A MeV/nucleon. 
We believe that the FOPI criterion is more sophisticated and corresponds better 
to the thermalization condition for the one-particle distribution functions. 
However, we should keep in mind the possible deviations caused by the event 
selection in the analysis of the data. 

\begin{figure}[tbh]
\includegraphics[width=8.5cm,height=13cm]{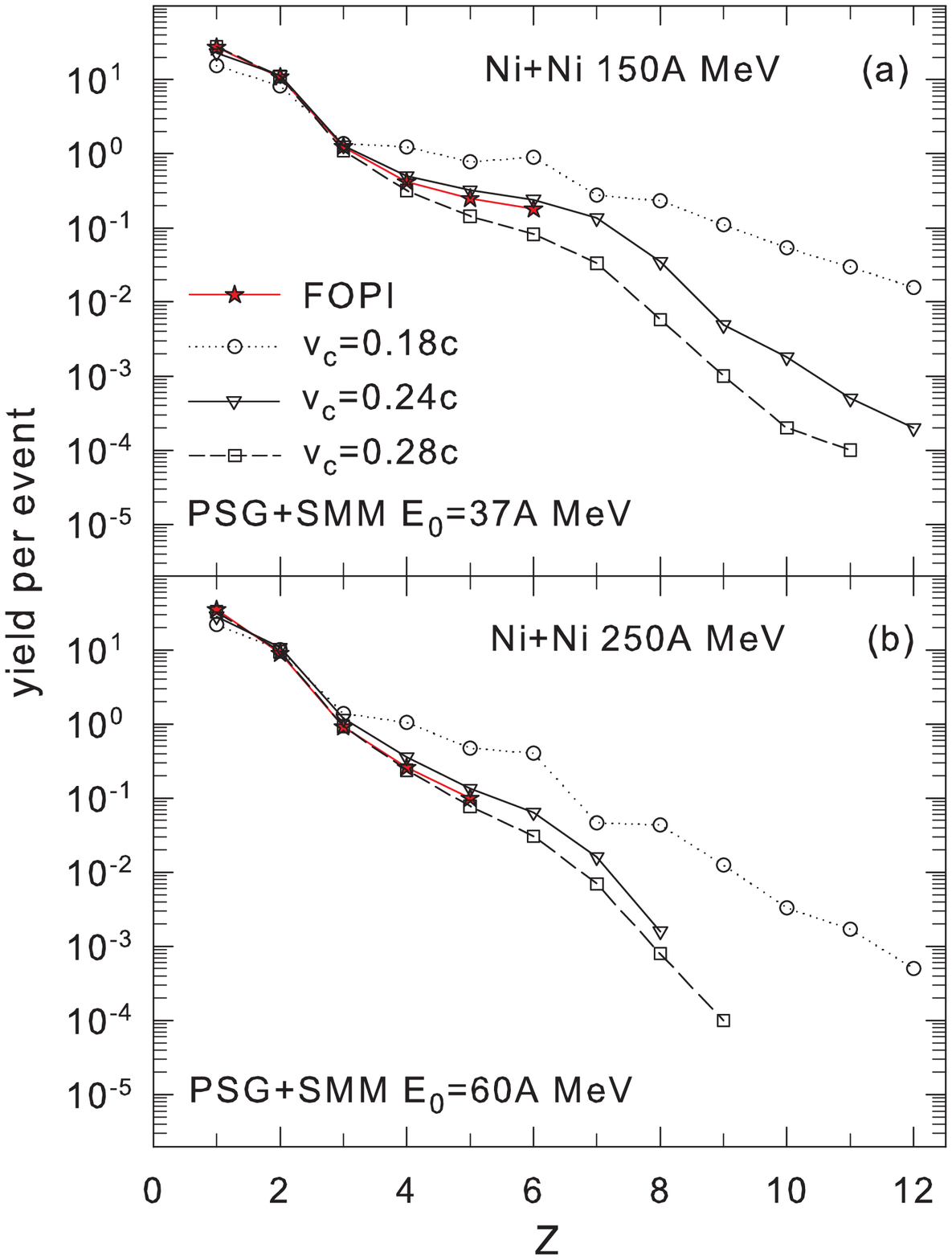}
\caption{\small{ 
Comparison of our calculations with the FOPI experimental data on the nuclei 
production in central Ni+Ni collisions at 150 A MeV (top panel (a)) and 
250 A MeV (bottom panel (b)). The parameters of the initial source are given in 
the figure. The nucleon distributions are after PSG, and parameters 
$v_c=$0.18$c$, 0.24$c$, and 0.28$c$ are used in the calculations. 
}}
\label{fig7}
\end{figure}

In Fig.~\ref{fig7} we show the comparison of our calculations of the charge 
yields of nuclei (which include the formation of primary nucleon clusters and 
their decay according to SMM) 
with the experimental data measured in Ni~+~Ni central 
collisions at 150 and 250 Mev per nucleon. We assume the formation of an 
initial system with $A_0$=116, $Z_0$=56, and total excitation energies of 
$E_{0}$=$37 A$ MeV and $E_{0}$=$60 A$ MeV, which correspond to the 
kinetic energy available in the center-of-mass 
(including the relativistic corrections). The proper energy/momentum 
balance was taken into account in the calculations. In order to show 
the dependence on the internal excitation energy and size of the primary 
coalescent-like clusters we present results for $v_c=$0.18$c$, 0.24$c$, and 
0.28$c$. We take the PSG method for the generation of the nucleons because 
it better suits to the FOPI event selection. 
One can observe a quite good agreement with the experimental data using 
the middle $v_c$, when the excitation of clusters are around 6--10 
MeV per nucleon. 

\begin{figure}[tbh]
\includegraphics[width=8.5cm,height=13cm]{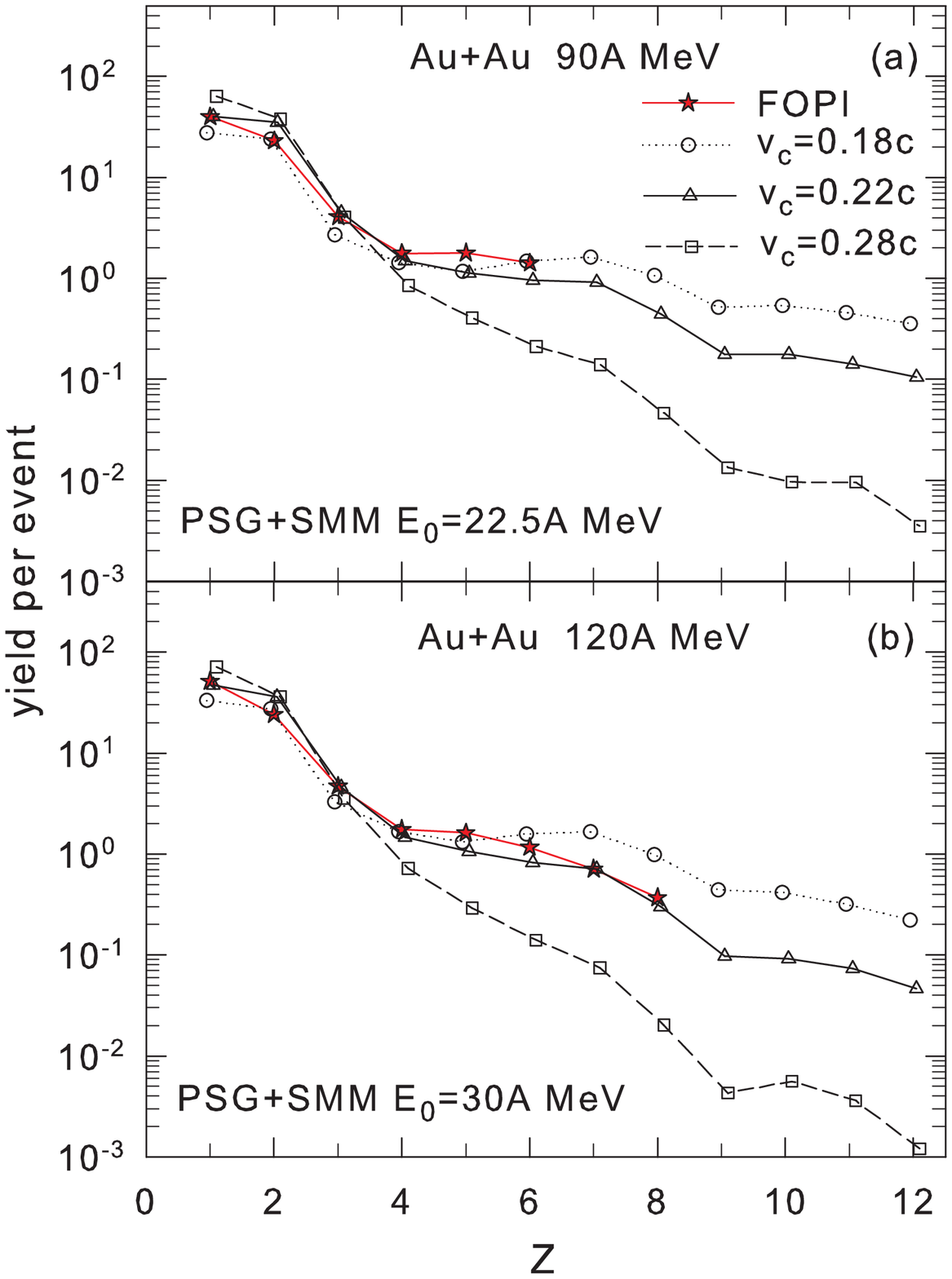}
\caption{\small{ 
Comparison of our calculations with the FOPI experimental data on the nuclei 
production in central Au+Au collisions at 90 A MeV (top panel (a)) and 
120 A MeV (bottom panel (b)). The parameters of the initial source are given in 
the figure. The nucleon distributions are after PSG, and parameters 
$v_c=$0.18$c$, 0.22$c$, and 0.28$c$ are used in the calculations. 
}}
\label{fig8}
\end{figure}

\begin{figure}[tbh]
\includegraphics[width=8.5cm,height=13cm]{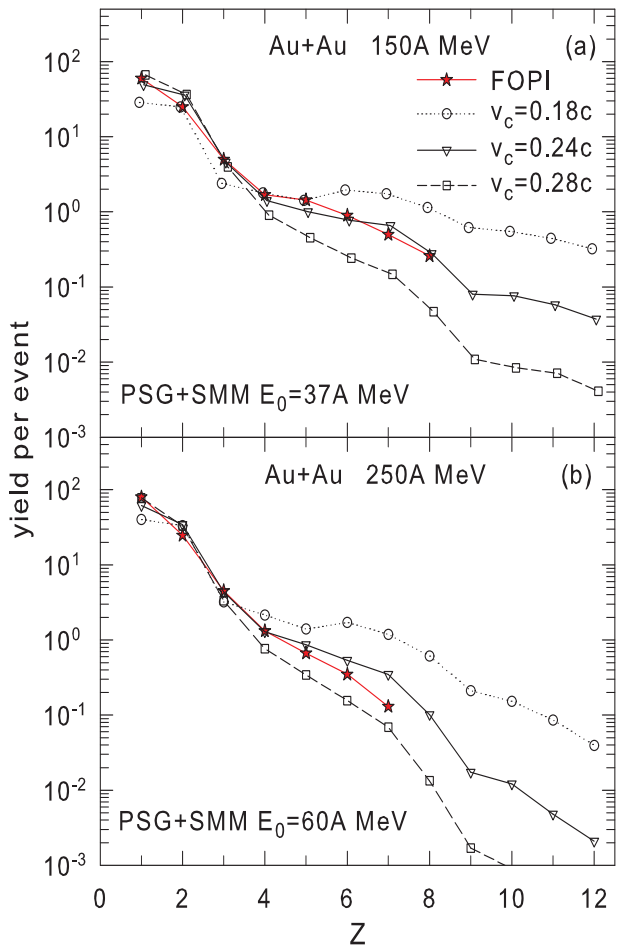}
\caption{\small{ 
The same as in Fig.~\ref{fig8}, but for central Au+Au collisions at 150 A MeV 
(top panel (a)) and 250 A MeV (bottom panel (b)). The  parameters $v_c=$0.18$c$, 
0.24$c$, and 0.28$c$ are used in the calculations. 
}}
\label{fig9}
\end{figure}

It is important to involve larger initial systems in the analysis. 
Fig.~\ref{fig8} and Fig.~\ref{fig9} present the comparison of our hybrid model 
calculations with the FOPI data on nuclei yields obtained in central Au+Au 
collisions at 90, 120, 150, and 250 A MeV (the corresponding center-of-mass 
energies are shown in the figures). 

\begin{figure}[tbh]
\includegraphics[width=8.5cm,height=8cm]{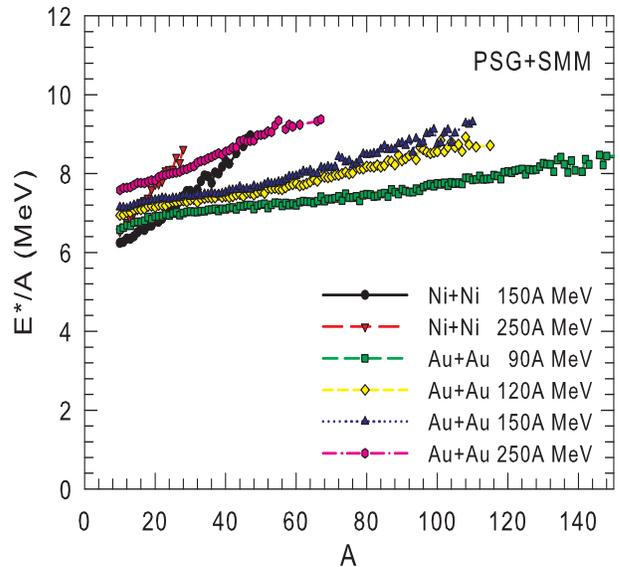}
\caption{\small{ 
Average excitation energy per nucleon ($E^{*}/A$) of local clusters of 
nuclear matter versus their mass number $A$ corresponding to the 
$v_c$ parameters which lead to the best description 
of the FOPI experimental data. The lines correspond to different reactions 
of central collisions, they are noted in the figure. 
}}
\label{fig10}
\end{figure}

One can also see for this selection a very good agreement with the experimental 
data. The small 
difference for the middle $v_c$ parameter is related to the fact that at 
low energies we need a slightly lower $v_c$ in order to construct the 
clusters with the appropriate internal excitation energy (6--10 MeV/nucleon), 
that 
is necessary for the successful description of the data. To illustrate 
this conclusion we provide in Fig.~\ref{fig10} the average excitation energy 
per nucleon($E^{*}/A$) distributions for the equilibrated clusters which give 
the best description of the data. By comparing with Fig.~\ref{fig3} one can 
see that addressing $E^{*}$ as a better parameter is quite justified since 
the analyzed range of $v_c$ can lead to very different values of 
$E^{*}$. Namely the excitation $E^{*}$, but not a $v_c$ 
parameter, has a physical meaning in our approach, since it gives defining 
information on the local equilibrated sources. As we know from the previous 
investigations of the statistical disintegration of excited finite nuclear 
systems (in multifragmentation reactions) these $E^{*}/A$ values correspond 
approximately to the temperatures $T\approx$6--8 MeV during the nuclear 
liquid-gas 
type phase transition in the phase co-existence region \cite{SMM,Ogu11}. 
One can see also that at low beam energy we obtain a rather massive clusters. 
Though, at high beam energy the cluster sizes become smaller, their excitation 
does not change. 
This saturation of the cluster excitation energy gives evidence that we are 
dealing with the local equilibrium phenomena. 

\begin{figure}[tbh]
\includegraphics[width=8.5cm,height=13cm]{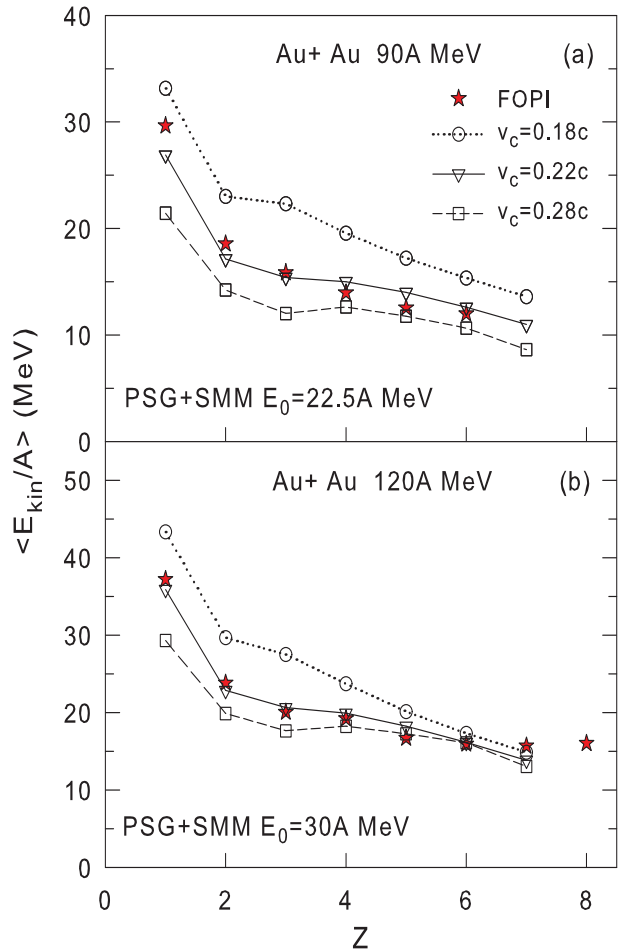}
\caption{\small{ 
Comparison of average kinetic energies per nucleon of produced 
nuclei versus their charges $Z$ with the FOPI experimental data in Au+Au 
central collisions. Top panel (a) is for 90 A MeV collisions, and bottom panel (b) 
is for 120 A MeV.
The parameters of the initial source are $A_0=$394, $Z_0=$158, and the energy 
$E_0$ is given in the figure. The nucleon distributions are after PSG. The 
parameters $v_c=$0.18$c$, 0.22$c$, and 0.28$c$ are used in the calculations. 
}}
\label{fig11}
\end{figure}

\begin{figure}[tbh]
\includegraphics[width=8.5cm,height=13cm]{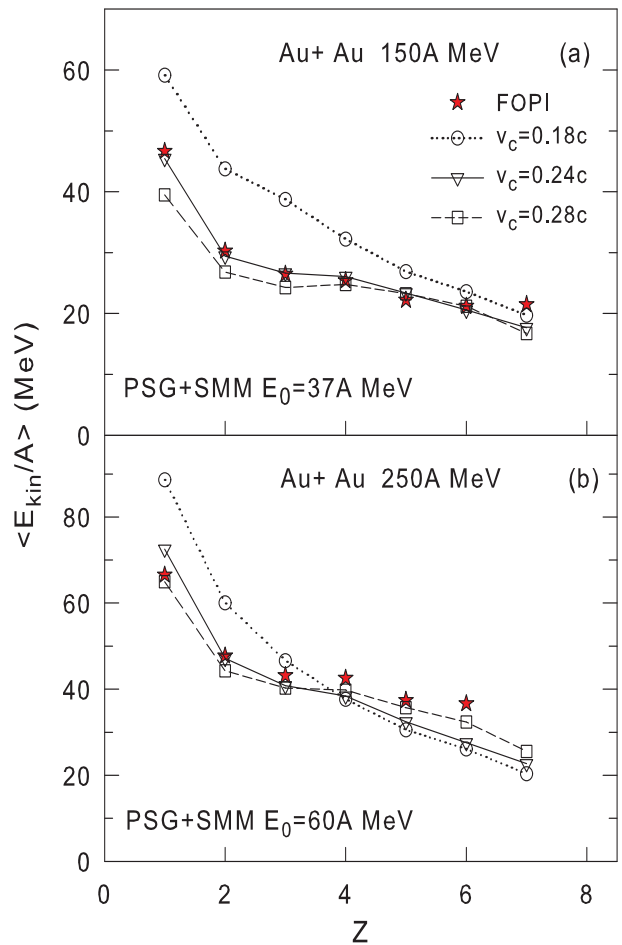}
\caption{\small{ 
The same as in Fig.~\ref{fig11}, but for central Au+Au collisions at 150 A MeV 
(top panel (a)) and 250 A MeV (bottom panel (b)). The  parameters $v_c=$0.18$c$, 
0.24$c$, and 0.28$c$ are used in the calculations. 
}}
\label{fig12}
\end{figure}

Another important experimental observable is the kinetic energy of the 
produced nuclei. Since the nuclei are formed from the nucleons belonging to 
the local clusters this kinetic energy depends on the initial energy of 
nucleons after the dynamical stage. When there is a correlation between the 
size of clusters and their positions in the expanding nuclear system we 
may infer that the correct description of the nuclei yields will lead also 
to their correct kinetic energies. In Fig.~\ref{fig11} and Fig.~\ref{fig12} we 
compare with the experimental data the kinetic energies of the nuclei per 
nucleon ($E_{kin}/A$) after their production. Sometimes this characteristic 
is associated with the flow energy. For the comparison we have selected the 
c.m. energies in the forward direction since they are better covered by 
the FOPI acceptance \cite{FOPI10}. We have taken the same calculations as 
for the yields presented in Figs.~\ref{fig8} and \ref{fig9}. One can see that 
the agreement is quite satisfactory for the best $v_c$ parameters (middle 
lines in the figures). The 
results on yields and energies are correlated with each other and 
support our conclusion. 

One can see  in Fig.~\ref{fig12} a slight difference between the predicted 
trends and 
the experimental data for the kinetic energies of the largest nuclei at the 
Au+Au 
collisions of the highest energy. As discussed already in Ref.~\cite{Bot21} the 
reason could be a slightly different initial energy distribution of nucleons 
than the PSG one. 
For example, with increasing beam energy the HYG distribution can contribute 
more to the ERAT event selection. This would automatically lead to high 
kinetic energies of large fragments (see  Fig.~\ref{fig6} and Ref.~\cite{Bot21}). 
However, as we have verified, 
the HYG nucleon distribution does not change our conclusion on the mechanism 
of the nuclei formation and on the internal excitation of primary clusters of 
nuclear matter under the equilibrium respective to the nucleation process.

\section{Discussion of the results} 

We have systematically analyzed the experimental production of light and 
intermediate mass nuclei obtained in central nucleus-nucleus collisions 
at high energy. We found that complicated many-body processes are responsible 
for the production of the nuclei. Specifically, it is not possible to describe 
them with one leading reaction 
channel. There are many transport models available, however, many of them are 
not equipped to treat the low-energy interaction of nucleons responsible for 
the nuclei formation in sufficient details. Because usual, they do not involve 
sufficiently realistic wave functions of nucleons, neglect details of 
many-body forces, collective interactions, and other processes 
important in this case. We have suggested 
a hybrid approach including dynamical and statistical reaction stages. 
Special attention is paid to the statistical description, and its 
generalization for the highly excited expanding nuclear systems. 
Since we believe that the formation of nuclei from nucleons can naturally 
take place at low nuclear densities in the later stage of the reaction process. 

Two phenomenological methods (PSG and HYG) are used to simulate the 
dynamical part of these high-energy reactions. 
These methods lead to quite different 
energy distributions of nucleons covering the most important limits expected 
after the initial dynamical stage. This stage is mostly determined by 
the high-energy interaction of individual nucleons. The nucleon system 
expands and low-energy interactions between neighbor nucleons result in the 
nuclei formation. Our hypothesis is that the nuclei are produced from the 
nucleons at a low density state of these expanding nuclear systems. If some 
dense nuclear clusters would be formed at large densities of nuclear matter, 
they would be destroyed by the subsequent interaction with other nuclear 
species during the expansion. Only when the system is sufficiently diluted one 
can expect the production of final nuclei. This situation is typical for the 
statistical freeze-out state. Therefore, we expect that the statistical 
approach should effectively work in this case. However, we are dealing with 
the finite nuclear systems. For this reason we must take into account that 
the interaction within such systems should be sufficient to apply the 
statistical laws. Usually, it is related to the excitation energy accumulated 
in the sources. We obtain from our analysis that one can use the 
statistical models at moderate excitations of the nuclear sources, around
maximum 6--10 MeV per nucleon, to describe the data obtained in central 
collisions. There will be several such sources (we construct 
them as coalescent-like clusters) in one highly-excited expanding nuclear 
system. This is an essential difference from the previous statistical 
description which has considered in reactions only one a such source. 
We suggest that local equilibrium is reached with a limited temperature 
respective to the 
nuclei formation in pieces (clusters) of expanding nuclear matter. One can 
consider it as a generalization of the statistical methods for nuclear 
systems formed in high energy reactions.

It is interesting that similar maximum excitation energies were extracted 
from the analyses of the multifragmentation data in relativistic peripheral 
nucleus collisions \cite{Bot95,Xi97,Ogu11}. However, it was done for single 
projectile-like sources remaining after the dynamical stage. In those 
collisions a source can expand to the freeze-out under thermal pressure or 
under the dynamical re-compression before the disintegration. Also low-excited 
compound nuclei are possible to produce from the projectile/target residues. 
Therefore, in those cases we have a very broad distribution of the source 
excitation energy from 0 up to $\sim$10 MeV per nucleon. The situation in 
central collisions is different. The whole nucleon system has already expanded 
after the dynamical stage and it can expand further entering the freeze-out 
state. It is concluded that nucleon clusters accumulating the excitations of 
6--10 MeV 
per nucleon respective to their ground states may be considered as statistical 
sources. They have a transition temperature ($T \approx 6-8$ MeV) and can be 
used for the statistical description of the fragment formation. 
In the considered finite systems with very high energy such 
a limitation of the temperature is related to selecting the clusters of smaller 
sizes. However, as is clear from our analysis, it will be inconsistent to 
involve very small clusters with low 
excitation energies. Since the interaction between nucleons leading to the 
fragmentation can already take place in the clusters of the intermediate size 
at the transition temperature.  

The obtained excitation energy (6--10 MeV per nucleon) is surprisingly 
close to the binding energy of the corresponding nuclear clusters. We 
assume that the binding energy can serve a natural energy to 
characterize a collective interaction necessary for the application of the 
statistical theory in finite systems. Also these 
excitations are sufficiently high in order to most of nuclear fragment 
formation processes in the clusters be well above their threshold. Therefore, 
the phase space can dominate over other kinematic restrictions for these 
processes and it can determine the final formation of nuclei. The statistical 
models are very effective for such many-body processes. 
On the other hand, when the clusters excitation energy is much higher 
than the binding energy we may expect that the one-particle scattering 
dominates and we can not effectively consider the nuclei formation. 

It is important for the statistical description, to confirm the 
local equilibration in finite expanding systems. 
The best experimental confirmation of these phenomena would be to measure 
particle correlations coming from the decay of the primary clusters during 
the nuclei production. Such correlations can bring direct evidences of the 
many-body character of the fragmentation process and the phase coexistence. 
Following our results, we predict the decreasing of the size of the local 
equlibrated clusters of nuclear matter with increasing total energy in the 
system, in order for the temperatures of the clusters not to change. This 
could clearly be seen in the correlations. Also, the differences from other 
nuclei formation mechanisms can be easily determined. For example, if we 
try to describe the cluster decay with the help of a transport 
model, which includes only one-particle distribution 
functions, in the end we can obtain many free nucleons and a large residue. 
While by applying the statistical model we obtain a lot of small nuclei 
($^2$H, $^3$H, $^3$He, $^4$He) in addition to a large residue. These different 
physical products can be observed in event-by-event correlation 
measurement. It is also important for the extension of the method to 
new phenomena: As was discussed in Ref.~\cite{Bot21} the correlation 
measurements would be specially instructive for the hypernuclei production 
to investigate the appropriate channels of their yield. 

It is instructive to emphasize once more the theoretical difference of our 
approach from the standard coalescence results. As in any phenomenology the 
coalescence parameter extracted from the comparison with experiment may depend 
on the initial stage description, and, therefore, it is entangled 
with this description. For example, the coalescence 
parameters for light nuclei obtained after the integration of nucleon spectra 
over all events \cite{Gos77,Som19} and the ones extracted from the 
event-by-event analysis \cite{Ton83,Bot17a} may be different by a factor 2--3. 
In our case we avoid 
this uncertainty. Independent of the dynamical stage the statistical clusters 
must have the excitation energy dependent on the nuclear liquid-gas phase 
transition properties. It has a clear physical meaning related to the nuclear 
matter. 

According to our analysis we can give a practical recipe for the calculation 
of the nuclei production in high energy nucleus collisions. This hybrid 
approach consists of several steps:  
1) The calculation of the nucleon distributions after the dynamical stage. 
One can use transport models, e.g., see 
Refs.~\cite{Bot21,Bot15,Bot17a,Som19,Ble99}. 
2) The selection of nucleon clusters at the low density of nuclear matter 
($\sim 0.1-0.3\rho_0$) and calculating their internal excitation energy. 
This should be done step by step, starting from large clusters and 
decreasing the cluster's sizes: The excitation energy decreases with the 
number of nucleons in the clusters. 
3) When the excitation energy is around 6--10 MeV per nucleon 
(close to the cluster binding energy) one can apply a statistical model to 
describe the cluster decay leading to the nuclei formation at the low 
density matter. 
We think that the uncertainty related to the cluster size will be small 
if their excitation is within the suggested range. Since at so 
high excitations the statistical models (see, e.g., \cite{SMM,Gross,Buy05}) 
lead to the 
scaling properties in the nuclei yields respective to the source size.

\section{Conclusions}

In the previous years the analyses of experimental data on disintegration 
of excited nuclear systems into nuclei 
\cite{SMM,ALADIN,Bot92,MMMC,Bot95,Xi97,Ogu11,EOS,Vio01,Pie02,FASA,
MSU,INDRA,Dag2} 
result in the conclusion that such fragmentation is of the statistical 
nature in many reactions. Also it was discussed that these processes can be 
the manifestation 
of the liquid-gas type phase transition in finite nuclei systems \cite{SMM}. 
As was obtained in the theoretical analyses of data on multifragmentation 
of relativistic projectiles \cite{Bot95,Xi97,Ogu11,Bot92,MMMC} 
there is an upper limit for the excitation energy for finite thermalized 
nuclear systems, around 10 MeV per nucleon, with the values close to the 
binding energies of normal nuclear systems. These systems decay in time 
about $\sim$100 fm/c \cite{Vio01,Pie02,FASA} after the beginning of the reaction 
that is several times longer than the initial dynamical reaction stage. 
We believe that it is a general property of finite nuclear systems: 
Independent on the way how the primary excited systems are formed, they 
can manifest the same properties of interacting nucleons in the region 
of the nuclear-liquid gas coexistence. 

In the present work we extend the statistical approach by considering 
the fragment production in central high energy nucleus collisions. 
We demonstrate that after the initial dynamical stage we can separate 
in each collision several excited statistical sources which decay 
producing the fragments. We call these small sources "coalescent-like" 
clusters, in order to emphasize the primary dynamics leading to the formation 
of such diluted nuclear systems. To simulate the dynamical stage we use 
the phenomenological phase space and hydrodynamical-inspired nucleon generators 
which provide the one-particle distributions, and which cover the most 
important limits of the nucleon momenta. The subsequent statistical decay 
of these clusters is the second part of our hybrid model. 
To verify our approach we have used the high quality FOPI data on the 
production of nuclei in central collisions \cite{FOPI10}.
Within our approach we have shown 
that it is possible to describe the fragment production with charges greater 
than one, including the yield and the kinetic energies, that was impossible 
with the previous methods. We believe it is also a result important for 
all statistical approach describing the disintegration of finite nuclear 
systems: Namely, the maximum excitation energy of such systems should be 
moderate, in the range of 6--10 MeV per nucleon. This is similar to the binding 
energies of the corresponding nuclei. Higher excitations are excluded 
as a result of our analysis. The lower excitations are possible in reactions 
of low energies at the formation of the single compound-like sources. However, 
in our case of high energy collisions the above mentioned cluster 
excitation provides the best description of the data. A limited temperature 
of the clusters is naturally caused by decreasing their sizes when more 
energy is deposited in the initial system during the nucleus collision. 
Actually, such clusters are locally equilibrated sub-systems within a 
very excited expanding nuclear system. In this respect the evolution of the 
nucleation process toward a high energy consists of a natural 
fragmentation of a low-density matter into such clusters. The statistical 
models are suitable for the decay of the clusters into nuclei at these energies 
because the nuclei production is essentially a many-body process and the phase 
space dominates in the nuclei formation process. 

Also we have pointed out that the correlations of the produced 
particles can be an important consequence of this kind of the fragment 
formation. We believe that this approach, with adequate dynamical 
models for the first reaction stage, should be used in future at high energies. 
It will give us a possibility to analyze new nuclear species formed from 
various baryons, e.g., hypernuclei \cite{Bot21}, which can be abundantly 
produced in central collisions. 

The authors acknowledges German Academic Exchange Service (DAAD) 
support from a PPP exchange grant and the Scientific and 
Technological Research Council of Turkey (TUBITAK) support 
under Project No. 121N420.  
This publication is part of a project that has received funding from the 
European Union~s Horizon 2020 research and innovation programme 
under grant agreement STRONG -- 2020 -- No.824093. 


\end{document}